# Polaron Coherence Condensation as the Mechanism for Colossal Magnetoresistance in Layered Manganites

N. Mannella[1,2,*], W. L. Yang[1,2], K. Tanaka[1,2], X. J. Zhou[1,2], H. Zheng[3], J. F. Mitchell[3], J. Zaanen[4], T. P. Devereaux[5], N. Nagaosa[6], Z. Hussain[2] and Z.-X. Shen[1,#]

[1]Departments of Physics, Applied Physics, Stanford University, Stanford, California 94305, USA
[2]Advanced Light Source, Lawrence Berkeley National Laboratory, Berkeley, California 94720, USA
[3]Materials Science Division, Argonne National Laboratory, Argonne, Illinois 60439, USA
[4]Instituut Lorentz for Theoretical Physics, Leiden University, POB 9506, 2300 RA Leiden, The Netherlands
[5]Department of Physics, University of Waterloo, Waterloo, Ontario, Canada N2L 3G1
[6]CREST, Department of Applied Physics, University of Tokyo, Bunkyo-ku, Tokyo 113, Japan

**ABSTRACT**

Angle-resolved photoemission spectroscopy data for the bilayer manganite $La_{1.2}Sr_{1.8}Mn_2O_7$ show that, upon lowering the temperature below the Curie point, a coherent polaronic metallic groundstate emerges very rapidly with well defined quasiparticles which track remarkably well the electrical conductivity, consistent with macroscopic transport properties. Our data suggest that the mechanism leading to the insulator-to-metal transition in $La_{1.2}Sr_{1.8}Mn_2O_7$ can be regarded as a polaron coherence condensation process acting in concert with the Double Exchange interaction.

$La_{1.2}Sr_{1.8}Mn_2O_7$ (x = 0.4, LSMO) is a prototypical bilayer manganite which exhibits the colossal magnetoresistive (CMR) effect by exploiting a transition from a ferromagnetic metallic (FM) groundstate to a paramagnetic insulating (PI) state above the Curie temperature $T_C$ = 120 K. Scattering measurements have provided direct evidence for the presence in the PI state of small lattice polarons with pronounced short ranged correlations forming a polaron glass phase as temperature approaches $T_C$ from above [1,2,3]. The mechanism responsible for the melting of the polaron glass into the FM state below $T_C$ remains unclear. Experimental and theoretical evidence has already suggested that the FM phase in the manganites is not a conventional metal as predicted by the Double Exchange (DE) model [4], but rather a polaronic conductor [5,6,7]. In the particular case of LSMO, recent Angle Resolved Photoemission (ARPES) experiment revealed that the FM phase is a polaronic metal with a strong anisotropic character of the electronic excitations, strikingly similar to the pseudogap phase in heavily underdoped cuprate high temperature superconductors (HTSC) [8]. Despite a strong mass enhancement and a small renormalization factor or pole strength $Z \approx 0.1$, the small but well-defined quasiparticle (QP) along the [110] or "nodal" direction is found to account for the metallic properties [8,9].

In this Letter, we present detailed ARPES measurements of the temperature dependent evolution of the QP excitations across the PI-to-FM transition. The coherent, polaronic FM groundstate emerges with well defined QPs which, despite a very small pole strength $Z$, track remarkably well the DC conductivity, consistent with the macroscopic transport properties. The onset of the FM groundstate is attained in a collective fashion by a kinetic energy-driven polaron condensation that dramatically enhances the spin DE interaction. This coherence-driven insulator-to-metal transition exhibits profound analogies to the case of underdoped HTSC, where the superconducting transition is controlled by the phase coherence of the Cooper pairs.

High-quality single crystals were grown with the floating zone method [10]. The ARPES measurements were carried out on beamline 10.0.1 at the Berkeley Advanced Light Source using a Scienta R4000 electron analyzer in the 14° angle mode, , with an angular resolutions of ± 0.15° ( ≈ 8.3 × $10^{-3}$ Å$^{-1}$ in momentum space). The samples were cleaved in ultrahigh vacuum (ca. 1-2 x $10^{-11}$ torr) at 30 K and measured with a total instrumental energy resolution of 20 meV.

Fig. 1 shows two Energy Distribution Curves (EDC) data sets collected below and above $T_C$. The polaron spectral function consists of a low energy QP peak representing the phase coherence of the dressed electrons, and a high energy incoherent peak (the hump, H) which tracks closely the bare dispersion, in essence reflecting the rapid motion of undressed carriers in a frozen lattice configuration [11]. Upon increasing temperature above $T_C$, the QP peak disappears as a result of the loss of coherence [8. On the contrary, the high energy branch does not show dramatic changes, although we note a shift towards higher binding energies (BE) of the hump peak position in the spectra corresponding to momenta $k$ approaching the Fermi momentum $k_F$, a phenomenology that we discuss in more detail below.

The temperature dependent evolution of the QP peak at $k_F$ is rather abrupt close to $T_C$, showing a collapse when the temperature is raised from T = 86 K to T = 124 K, a behaviour which can not be accounted by thermal broadening effects, but rather suggests a relationship to the macroscopic properties of LSMO (Fig. 2a). Surprisingly, the temperature dependence of the integrated spectral weight of the QP peak at $k_F$, which is directly proportional to $Z$ [12], tracks extremely well the DC conductivity curve $\sigma_{DC}$ as measured on samples grown in the same laboratory [13], thus yielding an empirical "$\sigma_{DC} \propto Z$" law (Fig. 2b). It is indeed remarkable that there exist such a net correspondence between DC conductivity data and an angle-resolved single particle excitation spectrum. To our knowledge, there is no theory of transport which can provide an explanation as to why $\sigma_{DC}$ should be entirely set by the QP pole strength $Z$. [14]. On the contrary, our data suggest a profound link between transport properties and QP coherence.

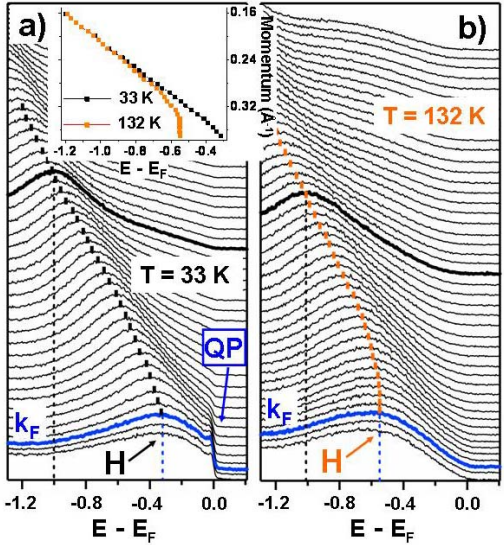

Fig. 1 (color online). Comparison between the $e_g$ band dispersions (a) below and (b) above $T_C$. The data have been collected along the [110] direction ((0,0) to ($\pi,\pi$), "nodal") with photon energy $h\nu = 42$ eV and light polarization vector laying in the sample plane. The blue lines denote the spectra corresponding to $k_F \approx 0.37$ Å$^{-1}$, while the spectra highlighted in black correspond to $k \approx 0.20$ Å$^{-1}$. The inset shows the dispersions of the hump peak positions.

This state of affairs resembles closely the temperature dependence of the QP intensity in the underdoped HTSC Bi$_2$Sr$_2$CaCu$_2$O$_{8+d}$ (Bi2212) at ($\pi$,0) [15]. Rather than at T*, the temperature at which the gap opens in the underdoped regime, the QP peak in underdoped Bi2212 exhibits an abrupt behavior at the superconducting transition temperature at which the phase coherence of the Cooper pairs sets in. Remarkably, the temperature dependence of the QP peak shows a resemblance to collective properties related to the superfluid density, which characterizes the phase coherence of the Cooper pairs [15,16].

The data here reported for LSMO and the analogies with the case of Bi2212 raise an intriguing and fundamental question: is there a common origin for the unexpected manifestation of collective effects in the single-particle excitation spectra in systems as different as manganites and underdoped HTSC? Our findings indicate that the commonality between LSMO and underdoped Bi2212 has to be identified in the coherence-driven nature of the MIT and superconducting transitions in LSMO and HTSC, respectively. While in HTSC the transition temperature $T_C$ denotes the onset of phase coherence of the Cooper pairs, in LSMO it marks the crossover from a low temperature coherent metallic state to a high temperature PI phase with diffusive, incoherent motion of small polarons.

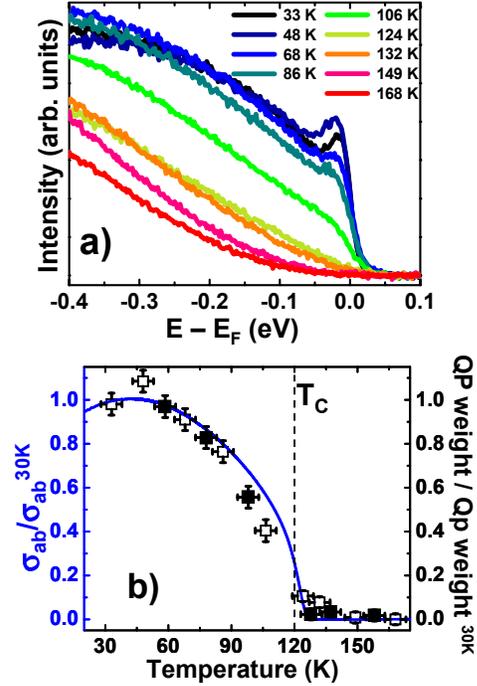

Fig. 2 (color online). (a) Temperature dependences of the QP peak at $k_F$ along the (0,0) to ($\pi,\pi$) "nodal" direction. (b) Integrated QP spectral weight and in-plane DC conductivity $\sigma_{ab}$ (from13). Both the QP weight and $\sigma_{ab}$ have been normalized to their value at 30 K. Open and filled squares denote warm up and cooling down cycles, respectively.

In general, the transition between the coherent and hopping polaronic regimes is expected to take place at a temperature of the order of some fraction of $\hbar\omega_0/k_B$, where $\omega_0$ denotes a fundamental phonon frequency [17,18]. In the case of LSMO, however, a mechanism is clearly at work which links intimately the degree of electron localization to the onset of ferromagnetism since the coherent-to-incoherent polaron crossover takes place at the Curie temperature $T_C$. The mechanism leading to the PI-to-FM transition in LSMO can be regarded as a polaron coherence condensation process acting in concert with the DE interaction. As the temperature is lowered below $T_C$, a complex many body wavefunction is attained in a dynamical percolation process due to the overlap of the small polarons wavefunctions as a result of the increase of their De-Broglie thermal wavelengths, thus transforming the classical insulating liquid of small polarons into a quantum-coherent, degenerate metal. The driving force for this transition is the sudden reduction of kinetic energy induced by the synergy of polaron condensation and the DE interaction. This synergistic interaction is characterized by a regenerative aspect of carriers delocalization: the polaron condensation fosters metallic conductivity by considerably increasing the overlap of the polaron wavefunctions on adjacent sites, thus enhancing the DE interaction and leading to ferromagnetism, while the DE interaction in turn enhances the carriers mobility and accelerates the polaron coherence condensation. We stress that both the polaron condensation and DE interaction are necessary for the MIT in LSMO. While the DE alone would not be sufficient to drive the MIT, it stills plays a crucial role in tipping the balance between the competing tendencies towards localization and delocalization.

The "$\sigma_{DC} \propto Z$" relation can be qualitatively rationalized by considering the metallic state as yielded by a superposition of a coherent quantum state, which we refer to as polaron groundstate, and a highly degenerate classical state at higher energy formed by small Jahn-Teller (JT) polarons, which we refer to as small polaron state. While the polaron groundstate is unique, the small polaron state is highly degenerate because the number of possibilities of forming a small JT polaron is of the order of the number of lattice sites. As the temperature is increased, the polaron groundstate is easily excited to the small polaron state with larger entropy. The fraction of polaron groundstate is thus a decreasing function of temperature, like $\sigma_{DC}$ in the metallic state, with the loss of polaron quantum coherence being signalled by the progressive disappearance of the QP peak.

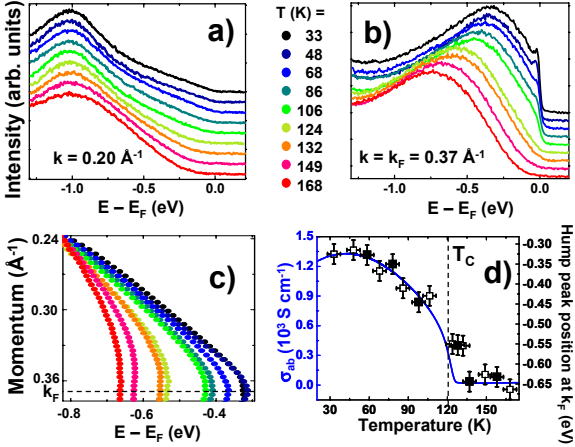

Fig. 3 (color online). Temperature dependence of the hump peak corresponding to (a) $k \approx 0.20$ Å$^{-1}$ and (b) $k_F \approx 0.37$ Å$^{-1}$. (c) Dispersions of the hump peak position at $k_F$ as a function of temperature. (d) Comparison between the variation with temperature of the hump peak position at $k_F$ and $\sigma_{ab}$ (from 13). Open and filled squares denote warm up and cooling down cycles, respectively.

We emphasize that by "polaron groundstate" we designate a distinctive state which is not to be identified with the canonical large polaron, but rather a state wherein all the particles in the system are locked in a complicated many body wavefunction whose expansion on a basis of single particle states yields the QP pole strength $Z \approx 0.10$. This estimate of $Z$ is in good agreement with the value determined from the ratio of the QP effective mass $m^*$ and the bare band mass $m_{LDA}$, namely $Z = m_{LDA}/m^* \approx 0.16$, a legitimate estimate when a possible momentum dependence of the self energy is neglected [19]. However, this direct correspondence between the value of $Z$ and the ratio $m_{LDA}/m^*$ is far from being trivial [20]. It indeed suggests that the polarons in the FM phase are almost on-site and not extended to the nearest neighbor sites. It is the overlap of the small polarons wavefunctions resulting from the increase of their De-Broglie thermal wavelengths that provides the coherence of the polaronic metal as the temperature is lowered.

Further insights about the peculiar nature of the polaronic metal are provided by the temperature dependence of the hump peak position. As already indicated in Fig. 1, while the spectra corresponding to momenta lower than $k \approx 0.24$ Å$^{-1}$ do not show any shift with temperature (Fig. 3a), the temperature dependence of the hump peak positions is very pronounced for the spectra at $k_F$ (Fig. 3b). The variation with temperature of the dispersions of the hump peak position is non-monotonic, and exhibits an abrupt shift close to $T_C$ for the spectra at $k_F$ (Fig. 3c), suggesting a relationship of this phenomenology to the macroscopic properties of LSMO. Surprisingly, the temperature dependence of the hump peak position at $k_F$ tracks $\sigma_{DC}$ extremely well (Fig. 3d).

The temperature dependence of the hump peak position at $k_F$ can be compared with that of the incoherent absorption (polaron) peak as measured by optical conductivity measurements [21,22]. We stress how remarkable is the energy scale (hundreds of meV) at which the spectral weight changes due to modest temperature variations in both ARPES and optics experiments. The hump peak position in ARPES is commonly identified with the polaron BE, thus providing a rationale for the large energy scale involved in the phenomenology described above [23]. At high temperatures, as expected in the case of strong coupling limit where the overlap of the polaron wavefunctions can be neglected (small polarons), the incoherent absorption peak measured by optics is found at roughly twice the BE of the hump peak in ARPES, indicating that the latter denotes indeed the polaron BE $E_P$. At low temperatures, however, the ARPES hump peak and the incoherent absorption peak in optics are not any more related in a simple way, most likely as a result of polaron wavefunctions overlap, suggesting that a full understanding of the polaron metal poses a much more complex many body problem than that of an isolated electron interacting with the lattice (small polaron). More importantly, the correspondence between the temperature dependences of the hump peak position at $k_F$ and the QP spectral weight at $E_F$ indicates that $Z$ and $E_P$ are related by a linear dependence like $Z \propto E_P/\hbar\omega_0$, whereas conventional polaron theories predict a much stronger dependence of the form $Z \propto \exp(-E_P/\hbar\omega_0)$, thus providing another distinctive characteristic of the coherent polaronic metal [17]. Our data suggest that the low (QP) and the high (Hump) energy scale physics are related: the loss of polaron quantum coherence, signaled as a collapse of the QP peak at very low energy, involves major rearrangements on a much higher energy scale visible as a considerable transfer of spectral weight from the QP peak to higher BE resulting in the shift of the hump peak position close to $k_F$. This shift is possibly related to the fact that the small JT polarons exhibit larger degrees of lattice distortions and a larger number of phonons than in the coherent polaron state below $T_C$.

In summary, we provide experimental evidence that the temperature variation of $\sigma_{DC}$ in LSMO is intimately related to that of the coherent QP weight $Z$, a quantity in essence representing the coherent fraction of the polaronic metallic groundstate. Our data suggest that the mechanism leading to the insulator-to-metal transition in LSMO can be regarded as a polaron coherence condensation process acting in concert with the DE interaction. The driving force for this transition is the sudden onset of coherence and concomitant reduction of kinetic energy induced by the synergistic and regenerative interaction between the polarons condensation and the DE interaction. The analogy of this effect to the case of underdoped Bi2212 suggests that coherence-driven transitions are a common denominator for the high temperature quantum phenomena in transition metal oxides.

The work at the ALS is supported by the DOE Office of Basic Energy Science, Division of Material Science, under contracts DE-FG03-01ER45929-A001 and DE-AC03-765F00515. The work at Stanford is supported by NSF grant DMR-0304981 and ONR grant N00014-04-1-0048-P00002.

* NMannella@lbl.gov, # zxshen@stanford.edu

[1] L. Vasiliu-Doloc, S. Rosenkranz, R. Osborn, S. K. Sinha, J. W. Lynn, J. Mesot, O. H. Seeck, G. Preosti, A. J. Fedro, and J. F. Mitchell, Phys. Rev. Lett. **83**, 4393 (1999).
[2] B. J. Campbell, R. Osborn, D. N. Argyriou, L. Vasiliu-Doloc, J. F. Mitchell, S. K. Sinha, U. Ruett, C. D. Ling, Z. Islam, and J. W. Lynn, Phys. Rev. B **65**, 014427 (2001).
[3] D. N Argyriou, J. W. Lynn, R. Osborn, B. Campbell, J. F. Mitchell, U. Ruett, H. N. Bordallo, A. Wildes, and C. D. Ling, Phys. Rev. Lett. **89**, 036401 (2002).
[4] C. Zener, Phys. Rev. **82**, 403 (1951).
[5] A. S. Alexandrov and A.M. Bratkovsky, Phys. Rev. Lett. **82**, 141 (1999); A.S. Alexandrov, Guo-meng Zhao, H. Keller, B. Lorenz, Y. S. Wang, and C. W. Chu, Phys. Rev. B **64**, 140404(R) (2001); A.S. Alexandrov, A.M. Bratkovsky and V.V. Kabanov, Phys. Rev. Lett. **96**, 117003 (2006).
[6] G. Zhao, V. Smolyaninova, W. Prellier and H. Keller, Phys. Rev. Lett. **84**, 6086 (2000); G. Zhao, D. J. Kang, W. Prellier, M. Rajeswari, H. Keller, T. Venkatesan, and R. L. Greene, Phys. Rev. B **63**, 060402(R) (2000).
[7] V.M. Pereira, J.M. Lopes dos Santos, A.H. Catro Neto, Cond-mat/0505741
[8] N. Mannella, W. Yang, X. J. Zhou, H. Zheng, J. F. Mitchell, J. Zaanen, T. P. Devereaux, N. Nagaosa, Z. Hussain and Z.-X. Shen, Nature **438**, 474 (2005).
[9] We extracted the value $Z \approx 0.10$ in the FM groundstate by first considering the spectrum corresponding to $k_F \approx 0.37$ Å$^{-1}$ at 30 K and then fitting the QP peak with a resolution-broadened lorentzian curve. Z is the ratio of the areas of the lorentzian curve and the peak-dip-hump structure after subtracting its inelastic background.
[10] J. F. Mitchell, D. N. Argyriou, J. D. Jorgensen, D. G. Hinks, C. D. Potter, and S. D. Bader, Phys. Rev. B **55**, 63 (1997).
[11] V. Perebeinos and P.B. Allen, Phys. Rev. Lett. **85**, 5178 (2000).
[12] The QP weight has been integrated in a ± 30 meV window around $E_F$ after subtracting a constant offset. Reducing or increasing the window size up to ≈ 60 meV, a value comparable to the QP peak width, results in the same QP weight variation with temperature. Since the same temperature dependence is obtained when the QP weight is normalized with respect to the intensity of the entire peak-dip-hump structure after subtracting a Shirley background, the temperature dependence of the QP weight can unambiguously be identified with that of $Z$.
[13] Q.A. Li, K.E. Gray, J.F. Mitchell, Phys. Rev. B **63**, 024417 (2000).
[14] In a Fermi liquid the scattering rate and the QP mass are both renormalized, and these renormalizations cancel each other in the expression for $\sigma_{DC}$. See A.J. Millis and P.A. Lee, Phys. Rev. B **35**, 3394 (1987).
[15] D. L. Feng, D. H. Lu, K. M. Shen, C. Kim, H. Eisaki, A. Damascelli, R. Yoshizaki, J.-i. Shimoyama, K. Kishio, G. D. Gu, S. Oh, A. Andrus, J. O'Donnell, J. N. Eckstein, Z.-X. Shen, Science **289**, 277 (2000).
[16] H. Ding J. R. Engelbrecht, Z. Wang, J. C. Campuzano, S.-C. Wang, H.-B. Yang, R. Rogan, T. Takahashi, K. Kadowaki, and D. G. Hinkset, Phys. Rev. Lett. **87**, 227001 (2001).
[17] A. S. Alexandrov and S. N. Mott, *Polarons and Bipolarons,* World Scientific (1995).
[18] S. Fratini and S. Ciuchi, Phys. Rev. Lett. **91**, 256403 (2003).
[19] J. W.Negele and H. Orland, *Quantum many-particle systems*, Perseus Books, 1988.
[20] Z and the ratio $m_{LDA}/m^*$ are in general different factors: while Z is determined by the overlap of the many body wavefunctions with and without an electron, the effective mass is sensitive to the wavefunctions overlap with electrons occupying neighboring sites.
[21] T. Ishikawa, T. Kimura, T. Katsufuji, Y. Tokura, Phys. Rev. B **57**, R8079 (1998).
[22] Myung Whun Kim, H. J. Lee, B. J. Yang, K. H. Kim, Y. Moritomo, Jaejun Yu, and T. W. Noh, Phys. Rev. Lett. **98**, 187201 (2007).
[23] The hump peak in ARPES represents the envelope of the Poissonian distribution of the phonon sidebands. See for example A.S. Alexandrov and J. Ranninger, Phys. Rev. B **45**, 13109(R) (1992); S. Ciuchi, F. de Pasquale, S. Fratini, D. Feinberg, Phys. Rev. B **56**, 4494 (1997).